\documentclass[12pt,preprint]{aastex}

\slugcomment{\today}

\def\ie{{\it i.e.\,}}
\def\ea{{\it et al. \,}}
\def\be{\begin{equation}}
\def\ee{\end{equation}}
\def\eg{{\it e.g., \,}}

\begin{document}
\title{POLARIZATION OBSERVATIONS OF THE ANOMALOUS MICROWAVE EMISSION IN THE PERSEUS MOLECULAR COMPLEX WITH THE COSMOSOMAS
EXPERIMENT}

\author{E.S. Battistelli$^{1,2}$, R. Rebolo$^{1,3}$, J.A. Rubi\~no-Mart\'{\i}n$^{1}$, S.R. Hildebrandt$^{1}$, R.A. Watson$^{1,4}$, C. Guti\'errez$^{1}$,
and R.J. Hoyland$^{1}$}

\affil{$^{1}$Instituto de Astrof\'isica de Canarias, Calle V\'ia
L\'actea s/n, E-38205 La Laguna, Tenerife, Spain.}

\affil{$^{3}$Consejo Superior de Investigaciones Cient\'ificas,
Spain.}

\affil{$^{4}$Jodrell Bank Observatory, University of Manchester,
Macclesfield, Cheshire SK11 9DL, UK.}

\altaffiltext{2}{Current address: Department of Physics and
Astronomy, University of British Columbia, 6224 Agricultural Road,
Vancouver, BC V6T 1Z1, Canada; e-mail: elia@phas.ubc.ca.}

\begin{abstract}

The anomalous microwave emission detected in the Perseus molecular
complex by Watson \ea has been observed at 11~GHz through dual
orthogonal polarizations with the COSMOSOMAS experiment. Stokes U
and Q maps were obtained at a resolution of $\sim0^{\circ}.9$ for a
30$^{\circ} \times $30$^{\circ}$ region including the Perseus
molecular complex. A faint polarized emission has been measured; we
find $Q=-0.2 \% \pm1.0\%$, while $U=-3.4^{+1.8}_{-1.4}\%$ both at
the 95\% confidence level with a systematic uncertainty estimated to
be lower than 1\% determined from tests of the instrumental
performance using unpolarized sources in our map as null hypothesis.
The resulting total polarization level is $\Pi =
3.4^{+1.5}_{-1.9}\%$. These are the first constraints on the
polarization properties of an anomalous microwave emission source.
The low level of polarization seems to indicate that the particles
responsible for this emission in the Perseus molecular complex are
not significantly aligned in a common direction over the whole
region, as a consequence of either a high structural symmetry in the
emitting particle or a low-intensity magnetic field. Our weak
detection is fully consistent with predictions from electric dipole
emission and resonance relaxation at this frequency.

\end{abstract}


\section{Introduction}\label{par:intro}

Since the Cosmic Background Explorer (COBE) experiment first
detected dust correlated microwave emission in its maps (Kogut \ea
1996), a significant effort has been made by the scientific
community to understand the origin of this anomalous emission and
characterize its properties. Further statistical evidence has been
found in observations and analysis by Leitch \ea (1997), de
Oliveira-Costa \ea (1997, 1998, 1999, 2002, 2004), Mukherjee \ea
(2001, 2003), Lagache \ea (2003), Finkbeiner (2004), Finkbeiner \ea
(2004) and Davies \ea (2005). Still, the underlying mechanism for
this emission is a matter of discussion due to the lack of
measurements about its properties, and different models have been
proposed to explain its observed characteristics.

The COSMOSOMAS experiment\footnote{See
http://www.iac.es/project/cmb/cosmosomas.} of the Instituto de
Astrofisica de Canarias (IAC) is now able to provide frequency and
sky coverage needed to increase our understanding of the statistical
and physical properties of the anomalous microwave emission
(Fern\'andez-Cerezo \ea 2006; S. Hildebrandt \ea 2006, in
preparation). Watson \ea (2005, hereafter W05) have recently
presented a direct detection of rising-spectrum emission by
COSMOSOMAS in the Perseus molecular cloud (R.A.=55$^{\circ}$.4,
decl.=31$^{\circ}$.8; J2000) that is an order of magnitude higher
than what can be explained with standard Galactic mechanisms of
emission (\ie, free-free, synchrotron, and thermal dust) and which
cannot be explained with ultracompact H II regions or a
gigahertz-peaked source. In this Letter, we present polarization
measurements of this anomalous emission performed with the
COSMOSOMAS experiment. It is worth stressing that the W05 detection
refers to an extended region that appears diffuse even at the
COSMOSOMAS angular resolution ($\sim 1^{\circ}$) with a Gaussian
FWHM fit over the emitting region of $\sim 2^{\circ}$. Our
polarization measurements are characterized by the same angular
resolution.

\section{Anomalous microwave emission: interpretation and
 polarization properties}\label{par:interpretation}

Comparisons between anomalous-emission detections and H$\alpha$ maps
have already ruled out the possibility that the statistical and
direct observations are due to free-free emission from ionized gas
(see, \eg Draine \& Lazarian 1998a, 1998b). Ultracompact H II
regions have also been invoked by McCullough \& Chen (2002) to
explain the direct tentative detection obtained by Finkbeiner \ea
(2002) as the superposition of a compact, optically thick and an
extended, optically thin H II region, both being free-free emitters.
Bremsstrahlung emission is intrinsically unpolarized, although
polarization may occur by means of Thomson scattering when photons
are rescattered within an H II region. This may occur in optically
thick regions, where the level of polarization can be at most 10\%
\citep{kea98}. Bennett \ea (2003a, 2003b) and Hinshaw \ea (2006)
interpret the dust-correlated component as synchrotron emission with
a flatter spectral index. However, the results of the Tenerife
Experiment (\eg de Oliveira-Costa \ea 2004) and the analysis of
Lagache (2003) are not consistent with this interpretation. In any
case, polarization measurements are essential to probe this
hypothesis, because the synchrotron emission is expected to be
highly polarized.

Draine \& Lazarian (1998a, 1998b) have proposed that the anomalous
dust-correlated emission is due to electric dipole emission from
rapidly rotating small dust grains (\ie "spinning dust") in the
interstellar medium. They have delved into the details of the
mechanisms of excitation and damping and concluded that their
emission spectrum may fit well the observed signal and be
responsible for this anomalous emission. Lazarian \& Draine (2000)
continued the study of spinning dust grains in the presence of weak
magnetic fields and found that paramagnetic relaxation resonance, in
a domain where classical paramagnetic relaxation is suppressed, may
be efficient at producing an alignment of grains rotating faster
than 1~GHz. This may result in the presence of an observable
polarization as high as 5\% at 10~GHz that rapidly decreases at
higher frequencies, in the anomalous emission, depending on the
phase in the sky of the magnetic field. Still, the level of
alignment depends on factors such as the efficiency of spin-lattice
relaxation which is uncertain for very small grains ($<10^{-7}$cm).

Magnetic dipole emission from dust grains has also been found to be
of considerable interest at frequencies below 100~GHz. Draine \&
Lazarian (1999) calculated the spectra of different grain candidates
and found that ordinary paramagnetic grains exhibit emission spectra
that are noticeably different from the observed anomalous emission.
However, they proposed a model, adjusting the magnetic properties of
the emitting material, that involves strong magnetic material whose
emission may account, at least in part, for the observed anomalous
emission. For this model, ferromagnetic relaxation may efficiently
act to align dust grains and produce strongly frequency- and
shape-dependent polarized emission that could be as high as 30\% at
10~GHz. A key signature of polarized emission from a magnetic
dipole, with respect to an electric dipole, is the variation of its
direction with frequency.

We note that both electric and magnetic dipole emissions are model
dependent, making it difficult to effectively predict their
polarization properties. However, according to the model proposed by
Draine \& Lazarian (1999), in order to account for the observed
spectrum, a hypothetical material "X4" characterized by strong
magnetic grains has to be considered. Therefore, asymmetric grains
will be aligned even by a low-intensity magnetic field like the
Galactic field, resulting in both a higher polarization level and a
higher probability for alignment to occur with respect to electric
dipole emission.

Iglesias-Groth (2005, 2006) studied the rotation rates of
hydrogenated fullerenes and electric dipole emission in the
interstellar medium and found that the smallest of these molecules
could be the origin of the anomalous emission detected by W05 in the
Perseus molecular complex and by Casassus \ea (2006) in the dark
cloud LDN 1622. In addition, weak ferromagnetic properties exhibited
by some of these molecules may lead to a consistent alignment and
consequently to detectable polarization.

\section{COSMOSOMAS observations}\label{par:obs}

\subsection{The instrument}

The COSMOSOMAS experiment is located at the Teide Observatory at
2400m above see level in Tenerife (Spain). It comprises two
circular-scanning instruments and generates two daily maps at 11~GHz
(COSMO11 instrument) and three maps at 13, 15, and 16~GHz (COSMO15
instrument). The angular resolution is approximately 0$^{\circ}$.9.
The interested reader may refer to Gallegos \ea (2001) for a general
description of the COSMO15 instrument and the adopted observational
strategy. COSMO11 has a similar instrumental setup to that of
COSMO15. The optical configuration is identical (dimensions are
scaled with wavelength) with a rotating primary flat mirror and a
secondary parabolic mirror focusing the radiation into a
cryogenically cooled receiver. The detectors are low noise HEMT
amplifiers cooled to 20 K.

The main difference between the two experiments is that COSMO11 is
optimized for polarization measurements. An ortho-mode-transducer
(OMT) is used to separate two orthogonal polarizations after the
radiation collector and before the HEMT, which are followed by
further amplification stages. This configuration allows one to
sample the sky and perform measurements of two Stokes parameters at
a time: I and Q (or U, depending on the orientation in the sky of
the measured polarization planes). In order to completely
characterize the linear polarization of the measured emission, the
sampled polarization planes can be rotated by 45$^{\circ}$ by simply
rotating the front end of the experiment, allowing one to switch the
receiver sensitivity from Q to U (and vice versa). A detailed
description of the instrument will be presented in R. Hoyland \ea
(2006, in preparation).

\subsection{Data Analysis}

COSMO11 creates daily maps with full coverage in right ascension and
$\sim 20^{\circ}$ coverage in declination. Raw data are stacked in
scan collections in which the detected data, resulting from the
adopted circular scan strategy, are saved as a function of time.
Each scan is the result of an average over 30 cycles (equivalent to
30~s). Atmospheric emission is evident in the scans as a
pseudo-sinusoidal modulation due to changing air masses within the
circles covered on the sky by the instrument. This modulation has
been removed by calculating a template and subtracting it from our
scans. The template was calculated by averaging over 2 hr, masking
bright sources and the Galactic plane, and iterating the procedure.
The time over which the template was calculated was chosen by
considering the final signal-to-noise-ratio as a function of time.
This procedure allows the subtraction of slow day-night modulation,
including 1/f noise as well as other fixed systematics such as
optical pickup, which is in any event strongly reduced by
under-illuminating the primary mirror. We note that this data
processing differs from that adopted in W05, and by
Fern\'andez-Cerezo \ea (2006), and is aimed at preserving the large
scales of the emission (\ie, to extend the window function to lower
multipoles $l$). After the cleaning procedure, we proceed to the
projection of the scans into the sky map, using bright sources to
check for consistency in the pointing reconstruction. The final
pixelization is $1/3^{\circ}\times1/3^{\circ}$. After that we sum
the daily maps into cumulated maps. The main calibration is
performed with Cygnus A on a daily basis. Absolute flux is taken
from the model by Baars \ea (1977). The instrumental beam is also
calculated using the main calibrator for each day. The overall beam
is recalculated using the cumulated map, and afterward we
recalibrate all the maps with fixed FWHMs with Gaussian fits over
the calibration source.

In calculating the source emission, we take great care to account
for contamination that might be due to the Sun or the Moon in the
proximity of the target sources. We have adopted a conservative
attitude, removing from the analysis those data in which the Sun is
closer than $\sim$30$^{\circ}$ to the regions of interest. Our
conservative attitude originates from concern about possible (even
very low) asymmetric pickup resulting in a polarized signal. A
significant effort has been made to track the instrument's stability
using radio sources and H II region emission and polarization with
time. Among other sources, we have checked in the reconstructed map
the coordinates of 4C 39.25 and 3C 345 outside the Galactic plane,
and 3C 84 and IC 405 in the Galactic anticenter, close to the
Perseus region of interest. These sources have allowed us to test
possible effects arising from the distance to the calibration source
in our maps, and to check that possible misalignment effects in the
optical system of the instrument are not affecting the local
reconstruction of the microwave sky in the region of interest. Also,
the relatively close H II region NGC 1499 (the California Nebula)
has been used to test for possible effects and systematics arising
from measurements of extended sources compared with pointlike
sources. All of these sources have shown a good level of stability
or slow variation compatible with intrinsic flux variations. NGC
1499 has also been carefully studied to check for spurious effects
and as a null hypothesis test, since diffuse H II regions are
expected to mainly emit free-free radiation, which is intrinsically
unpolarized.

Possible diffuse polarized synchrotron emission has been checked
using the maps from Wolleben \ea (2006) \footnote{See
http://www.drao-ofr.hia-iha.nrc-cnrc.gc.ca/26msurvey/data.html.}.
Wolleben \ea point out the presence of depolarized regions around H
II regions. Fortunately, this does not have an effect on the
polarized emission of the Perseus molecular cloud, which is known to
be closer to us than NGC 1499. A detailed analysis of the Wolleben
\ea polarization maps, and the extrapolation toward COSMOSOMAS
frequencies, shows possible residual polarized signal lower than 1\%
in the NGC 1499 region, even assuming a spectral index as high as
-2.7. We have thus finally re-calibrated our Perseus emission with
NGC 1499.

The coordinate convention adopted in this Letter to define the Q and
U Stokes parameters is as follows. At every point on sky, the X-axis
points north, and the Y-axis points east. In the first COSMO11
configuration, we measure $I_{0^\circ}$ (intensity along the X-axis)
and $I_{90^\circ}$ (intensity along the Y-axis), while in the second
orientation of the system we obtain $I_{+45^\circ}$ and
$I_{-45^\circ}$. Our definition of the Stokes parameters is $Q =
I_{0^\circ} - I_{90^\circ}$ and $U=I_{45^\circ} - I_{-45^\circ}$.
For each of the two configurations, we have two determinations of
the intensity, which we denote $I_Q = I_{0^\circ} + I_{90^\circ}$
and $I_U=I_{45^\circ} + I_{-45^\circ}$. Finally, the total
polarization degree is defined as  $\Pi = \sqrt{Q^2+U^2}/I$.

Spurious instrumental polarization induced by oblique reflections in
our off-axis system has been calculated, in a regime of ordinary
skin effect, using the method presented by Renbarger \ea (1998), and
found to be totally negligible compared with other systematics. The
two channels have almost identical spectral response, resulting in a
negligible systematic effect arising from the passband mismatch
between observed calibrator and the Perseus region (\ie, the effect
is canceled out). Further instrumental effects have been tested for
by rotating the frontend of the instrument by 90$^{\circ}$ and
checking for consistency.

Measurements of Q were taken over the period between 2004 March and
2005 May, while measurements of U are extracted from the
measurements performed after this month. The time coverage is not
uniform, as a result of contaminations, variation in the atmospheric
conditions, and instrumental failures, which have been reduced to a
minimum thanks to continuous monitoring of the instrument's
performance. The integration time for the U-measurements is smaller
than that for Q, resulting in higher statistical errors. The
residual systematics are monitored by measurements of NGC 1499,
using this H II region as a null test. This is done in two steps: by
directly observing NGC 1499 on the Cygnus-calibrated map, and then
by considering the possibility of a faint polarization of the main
calibrator which could result in an apparent polarization signal of
the H II region. Tests on the Wolleben \ea (2006) maps allowed us to
check for this possibility and, after accounting for this, to
constrain the systematics for both  the Q- and U-directions at a
lower level than the statistical uncertainties. In order to
double-check this result we have carefully monitored 3C 84, whose
emission is expected to be variable with time but whose polarization
is expected to be far below 1\% \footnote{See the UMRAO database, at
http://www.astro.lsa.umich.edu.}. We find a final polarization level
lower than 1\%, which sets our systematics level.

\section{Results and discussion}\label{par:results}

The effective polarized emission from the Perseus region has been
calculated through a maximum likelihood analysis using the
measurements of the partial stacked maps. This likelihood was built
using a multivariate Gaussian distribution. Measurements at
different epochs were assumed to be uncorrelated, so the covariance
matrix in this case is diagonal. We describe the data with a
two-parameter model, one for the value (assumed to be constant in
time) for the polarization of the Perseus region, and another for
the calibration uncertainty. This latter parameter is marginalized
over by following the analytical prescription given by Bridle \ea
(2002), assuming a Gaussian prior for its value with $10\%$ width
(overall calibration uncertainty). Note that in the case of the
fractional polarization, the calibration uncertainty does not enter.
Once the likelihood curves are computed, the confidence limits are
derived from the 0.025, 0.5, and 0.975 points of the cumulative
distribution function. Thus, our parameter estimate is the median of
the marginalized posterior probability distribution function, and
the confidence interval encompasses 95\% of the probability. In
Figure 1, we present the map of the Perseus region emission and the
two difference maps of 90$^{\circ}$ polarization orientation
describing the Q and U parameters.

We find a small polarization in both orientations: for the
$0^{\circ}/90^{\circ}$ orientation we obtain $Q/I=-0.2 \% \pm1.0\%$,
while for the $45^{\circ}/-45^{\circ}$ one we have $U/I=-3.4
^{+1.8}_{-1.4}\%$ both at the 95\% confidence level, for a total
polarization of $\Pi = 3.4^{+1.5}_{-1.9}\%$. The polarization in the
$0^{\circ}/90^{\circ}$ orientation is thus consistent with a null
measurement within the systematic and statistical uncertainties. A
polarization in U is possible, both from a statistical and a
systematic point of view. From the considerations highlighted in the
\S1, although model-dependent uncertainties are widely present, the
low level of polarization in the Perseus anomalous emission appears
to be inconsistent with magnetic dipole emission of highly oriented
particles in the presence of a magnetic field, either because of low
field intensity in the region, or because of the highly symmetric
nature of the emitting particles. Our results favor electric dipole
emission as being responsible, the emission properties of which are
even more model dependent but whose polarization is limited to $\sim
5\%$. In particular, as described by Lazarian \& Draine (2000),
paramagnetic relaxation resonance may produce an observable
polarized signal of the same order as that observed by COSMOSOMAS
for an equivalent grain radius of $\lesssim 10^{-7} cm$. Our
observations are limited both by the interpretation of the models
and by our angular resolution. In fact, we cannot constrain
magnetically emitting grains characterized by weak magnetic
properties or by spherical shape, and we cannot monitor structures
characterized by angular dimensions smaller than our angular
resolution in the observed region. This latter problem is, however,
somewhat lessened by the fact that the measurement from W05 is over
an extended region and so we expect anomalous-emission properties to
characterize the entire region. Further polarization observations
with higher sensitivity and angular resolution, as well as
monitoring other frequencies of interest for the anomalous emission,
will further help in understanding the mechanism responsible for
this emission.

\section{Conclusions}

We have observed the Perseus molecular cloud in order to extract
information about the polarization properties of the anomalous
microwave emission observed by W05. A careful analysis and control
of systematics lead us to the conclusion that this emission is
characterized by a low level of polarization: $\Pi =
3.4^{+1.5}_{-1.9}\%$ at 95\% confidence level, with systematic
uncertainties limited to 1\%. To infer the real emission mechanism
for the anomalous emission, local physical properties of the
observed regions should be further investigated. However, the weak
detected polarization seems to support the electric dipole emission
model, with resonance relaxation, over the dipole magnetic emission
hypothesis. Further information about the effective origin of this
emission will be available with higher resolution, higher
sensitivity, multifrequency observations already planned at the
Teide Observatory and, hopefully, planned for other instruments.

\acknowledgments We acknowledge the support of the IAC and the staff
of Teide Observatory for the construction and operation of the
COSMOSOMAS experiment. Partial funding was provided by grant
AYA2001-1657 from the Spanish Ministry of Science and Education. We
thank M. Wolleben for the online 1.4 GHz maps. We thank E. Chapin
for useful comments and for revising the English. Also, we would
like to thank the referee for useful comments that allowed us to
improve this Letter.

\clearpage

\begin{figure}
\plotone{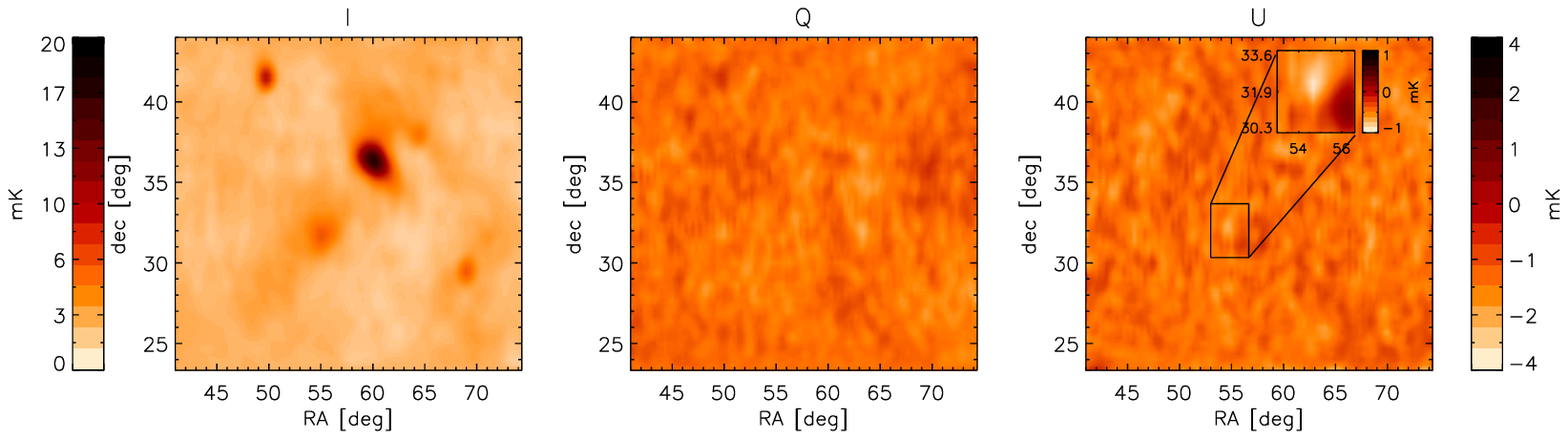} \caption{Stokes I, Q, and U COSMOSOMAS 11
GHz maps of the region around the Perseus molecular complex. In the
I map, the brightest source is the NGC 1499 H II region. The fainter
source at R.A.=55$^{\circ}$.4, decl.=31$^{\circ}$.8 (J2000) is the
observed (anomalously) emitting region. The source 3C 84 is also
visible in the intensity map, with coordinates R.A.=49$^{\circ}$.9
and decl.=41$^{\circ}$.5 (J2000). The color bar on the left refers
to the I map, while that on the right refers to both Q and U maps.
The maps have been smoothed with a 3 pixel boxcar. The inset in the
U map refers to the faint detection achieved in the Perseus
molecular complex.} \label{maps}
\end{figure}

\end{document}